\documentclass[prd,amsmath,amssymb,superscriptaddress,floatfix,nofootinbib,10pt]{revtex4}
\usepackage{times}
\usepackage{amssymb,amsbsy,amsmath,amsfonts}
\usepackage{graphicx}
\usepackage{float}
\usepackage{xcolor}
\usepackage{orcidlink}
\usepackage{morefloats}
\usepackage{rotating}
\usepackage{srcltx}
\usepackage{slashed}
\usepackage{multirow}
\usepackage{verbatim}
\usepackage{hyperref}
\usepackage{tabularx}
\usepackage{bm}
\allowdisplaybreaks[4]

\usepackage{bbding}
\usepackage{threeparttable}

\usepackage{mathrsfs}

\DeclareUnicodeCharacter{2212}{\ensuremath{-}}

\newcommand{\PreserveBackslash}[1]{\let\temp=\\#1\let\\=\temp}
\newcolumntype{C}[1]{>{\PreserveBackslash\centering}p{#1}}
\newcolumntype{R}[1]{>{\PreserveBackslash\raggedleft}p{#1}}
\newcolumntype{L}[1]{>{\PreserveBackslash\raggedright}p{#1}}

\def \R{\textcolor{black}}

\begin{document}

\title{Exotic multistrange-anticharm baryon systems}

\author{Jing Song\,\orcidlink{0000-0003-3789-7504}}
\email[]{Song-Jing@buaa.edu.cn}
\affiliation{School of Physics, Beihang University, Beijing, 102206, China}%
\affiliation{Departamento de Física Teórica and IFIC, Centro Mixto Universidad de Valencia-CSIC Institutos de Investigación de Paterna, 46071 Valencia, Spain}

\author{ Eulogio Oset\,\orcidlink{ https://orcid.org/0000-0002-4462-7919}}
\email[]{oset@ific.uv.es}
\affiliation{Departamento de Física Teórica and IFIC, Centro Mixto Universidad de Valencia-CSIC Institutos de Investigación de Paterna, 46071 Valencia, Spain}
\affiliation{Department of Physics, Guangxi Normal University, Guilin 541004, China}

\begin{abstract}
We study exotic baryon systems composed of anticharmed mesons and strange baryons using the extended local hidden gauge approach. By solving the coupled-channel Bethe-Salpeter equation with interaction kernels from vector meson exchange, we explore the formation of hadronic molecular states in sectors with strangeness $S = -1$, $-2$,  $-3$ and $-4$. 
\textcolor{black}{We systematically consider all possible isospin configurations and include both octet and decuplet baryons in the coupled-channel systems.}
\textcolor{black}{Our results indicate that attractive interactions in $S=-1,-2$ can dynamically generate bound states, while systems with $S=-3,-4$ have repulsive interactions and do not support molecular formation.}
We also investigate vector-baryon systems with $\bar{D}^*$ and $\bar{D}_s^*$ mesons, finding similar but more deeply bound states. 
The results show that bound exotic states are more likely when one or two strange quarks are present. 
\textcolor{black}{To assess the robustness of our predictions, we perform an uncertainty analysis by varying the cutoff parameter \( q_{\text{max}} \), which affects the loop function regularization. The variations lead to moderate shifts in the pole positions, confirming the qualitative stability of the molecular states.}
These results highlight the strangeness dependence of the molecular formation mechanism and provide theoretical predictions that can guide future experimental searches for exotic multistrange-anticharm baryon systems. 

\end{abstract}


\maketitle

\section{Introduction}
\label{intro}

The investigation of exotic hadronic states has been a central focus in hadron spectroscopy over the last two decades, especially following the discovery of the $X(3872)$ by the Belle collaboration \cite{Belle:2003nnu}. These studies have expanded our understanding of hadrons beyond the conventional three-quark baryons and quark-antiquark mesons, introducing the possibility of hadronic molecules and multiquark configurations~\cite{Jaffe:1976ig, GellMann:1964nj, Esposito:2016noz, Guo:2017jvc, Olsen:2017bmm, Lebed:2016hpi}. 
Among these exotic configurations, systems involving anticharm quarks and multiple strange quarks have attracted  attention due to their unique quantum numbers, exotic flavor content, and the opportunity that they provide  tests of  the dynamics of QCD in the non-perturbative regime, which attracts much attention \cite{Chen:2016qju, Liu:2019zoy, Brambilla:2019esw, Ali:2017jda, Karliner:2017qhf}. 

Recent results from the LHCb Collaboration report the first observation of the decay $\Lambda_b^0 \rightarrow D^+ D^- \Lambda$~\cite{LHCb:2024hfo}, revealing indications of intermediate resonant structures in the $D^+ \Lambda,~D^- \Lambda$ and $D^+ D^-$ invariant mass distributions. A baryon resonance decaying to $D^-\Lambda$ would be necessarily exotic, since it contains the $\bar cd sud$   quarks, and no  $q\bar q$  can annihilate.

Hadronic molecules, which are loosely bound states of mesons and baryons, have proven to be a useful framework for interpreting many of the newly observed exotic hadrons~\cite{Tornqvist:1993ng, Swanson:2006st, Guo:2017jvc, Chen:2016qju, Esposito:2016noz, Olsen:2017bmm, Lebed:2016hpi}. While many studies have focused on hidden-charm or open-charm systems with little or no strangeness, less attention has been paid to configurations that combine both anticharm and multiple strange quarks. These multistrange-anticharm systems could offer valuable insight into the underlying QCD forces and are not easily explained within conventional quark models.

Theoretically, meson-baryon interactions in the light quark sector involving  strange quarks can dynamically generate bound or resonant states through coupled-channel 
unitarization~\cite{Kaiser:1995eg,Oset:1997it,Oller:2000fj, Jido:2003cb,Hyodo:2011ur, Roca:2005nm, Garcia-Recio:2005elc, Gamermann:2007mu}. This approach uses effective Lagrangians consistent with chiral symmetry, combined with the Bethe-Salpeter equation in its on-shell factorized form, and has successfully explained many resonances, including the emblematic two $\Lambda(1405)$ states. Extrapolations to the heavy quark sector have also successfully explained a range of exotic states, including hidden-charm pentaquarks~\cite{Roca:2016tdh,Roca:2015dva,Wu:2010jy, Wu:2010vk, Xiao:2013yca}, doubly charmed baryons~\cite{Dias:2018qhp,Wang:2022aga,Ramos:2020bgs,Liu:2019zoy}, and hidden charm strange or open charm strange resonances~\cite{Montana:2018edp,Montana:2017kjw,Roca:2024nsi,Wang:2019nvm,Liu:2020hcv,Lin:2023iww,Liu:2018bkx,Wang:2018alb,Debastiani:2018adr,Huang:2018wgr,Debastiani:2017ewu}. 

Recent experimental discoveries, such as the $P_c$ pentaquarks~\cite{Aaij:2015tga, Aaij:2019vzc}, the $T_{cc}^+$ tetraquark~\cite{LHCb:2021vvq,LHCb:2021auc}, and the excited $\Omega_c$ states~\cite{LHCb:2017uwr}, show that the hadron spectrum is richer than previously expected. These narrow structures encourage further investigations into systems containing both anticharm and high strangeness. In such systems, the interplay between heavy quark symmetry (HQS) and SU(3) flavor symmetry  can lead to strong interactions, possibly resulting in bound states near meson-baryon thresholds. The interactions used in these studies are often derived from the extended Weinberg-Tomozawa (WT) term, which incorporates constraints from both chiral symmetry and HQS~\cite{Hofmann:2005sw,Mizutani:2006vq, Tolos:2007vh,Garcia-Recio:2008rjt,Romanets:2012hm}.

In this work, we extend this line of research to unexplored configurations involving anticharmed mesons and strange baryons. These systems are considered  as candidates for hadronic molecules that lie beyond the conventional baryon spectrum. Inspired by earlier successes in describing hidden-charm and open-charm pentaquarks and tetraquarks as hadronic molecules~\cite{Guo:2017jvc,Wu:2010jy, Wu:2010vk, Yang:2011wz, Xiao:2013yca, Xiao:2019gjd, Dias:2014pva, He:2019ify, Sakai:2019qph}, we explore whether similar dynamics can lead to bound or resonant states in the anticharm sector with increasing strangeness.

We consider meson-baryon systems consisting of an anticharmed meson ($\bar{D}$ or $\bar{D}_s$) and strange baryons ($\Lambda$, $\Sigma$, $\Xi$, $\Omega$). The coupled channels included in our study are:
$
(\bar{D}_s N,\quad \bar{D} \Lambda,\quad \bar{D} \Sigma),\quad (\bar{D}_s \Lambda,\quad \bar{D} \Xi),\quad \bar{D}_s \Xi,\quad \text{and} \quad \bar{D}_s \Omega,
$
and they are all of exotic nature. These channels allow us to explore systems with strangeness ranging from $-1$ to $-4$.

\R{We further distinguish these configurations by their isospin structure. For $S=-1$, we study channels such as $\bar{D}_s N$, $\bar{D} \Lambda$, and $\bar{D} \Sigma$ with isospin $I=1/2$, as well as $\bar{D} \Sigma$ with $I=3/2$. For $S=-2$, we include $\bar{D}_s \Lambda$ and $\bar{D} \Xi$ with $I=0$, along with $\bar{D} \Xi$ with $I=1$. For $S=-3$, the channel $\bar{D}_s \Xi$ is considered with $I=1/2$. These isospin-resolved configurations allow a detailed exploration of the possible bound or resonant states that can arise in different flavor combinations.}

\R{
In addition to the pseudoscalar anticharmed meson–baryon systems, we also extend our study to include interactions involving the corresponding vector mesons ($\bar{D}^*$, $\bar{D}_s^*$) and strange baryons. The vector-baryon channels span a broad range of strangeness values and provide additional opportunities for forming molecular states or resonances. The dynamics in these vector systems closely resemble those of the pseudoscalar cases, allowing us to explore a richer spectrum of exotic hadronic configurations.
}

{\color{black}In this work, we  include both baryon octet and baryon decuplet states in our coupled-channel analysis. Consideration of the decuplet baryons extends the range of possible molecular configurations and allows for a more complete understanding of the dynamics, as the spin and flavor structure of decuplet baryons can influence the interaction potentials and the formation of states.}

To study these systems, we solve the Bethe-Salpeter equation with interaction kernels derived from the extended WT term, which is most practically implemented using the local hidden gauge approach~\cite{Bando:1984ej,Bando:1987br,Meissner:1987ge,Nagahiro:2008cv}. These kernels respect both chiral symmetry, when the interaction involves light quarks, and heavy quark symmetry when the interaction involves heavy quarks. Similar approaches have been successfully applied to  related sectors, such as $\bar{D} N$, hidden charm and open charm interactions, which  have shown the possibility of forming bound or resonant states via $S-$wave interactions \cite{ Haidenbauer:2010ch, Yamaguchi:2011xb, Wang:2011rga,    Garcia-Recio:2011jcj,Xiao:2019gjd}. This motivates us to explore the multistrange-anticharm systems in a similar framework.

On the experimental side, facilities such as the LHC, Belle II, J-PARC, and BESIII provide promising environments for producing such exotic states in high-energy collisions~\cite{Brambilla:2019esw,Han:2024duu,Szabelski:2025wxk,BESIII:2020qkh,BESIII:2013ris}. Accurate theoretical predictions for the masses, decay widths, and dominant decay channels of these states will be crucial to guide future experiments.

There is some theoretical work done on these states. Early quark models favor the existence of bound pentaquarks of $\bar c sqqq$ nature~\cite{Lipkin:1987sk,Gignoux:1987cn}. From the molecular perspective there are works finding bound states in the $S=-1,-2$ states~\cite{Hofmann:2005sw,Yan:2023ttx,Yalikun:2021dpk} which we will discuss from the perspective of the results obtained here.

\section{Formalism}

\noindent We consider meson baryon coupled channels with a  $\bar{D}~(\bar{D}_s)$ mesons and baryons with different strangeness, we have,
\R{
\begin{itemize}
\item \textbf{Octet baryon}s:
\end{itemize}
}
\vspace{0.2cm}
\noindent $S=-1,~ \text{channels}: \quad \bar{D}_s N,~ \bar{D} \Lambda,~ \bar{D} \Sigma,\quad \text{with isospin}~~ I=1/2$;\quad \R{ $\bar{D} \Sigma,\quad \text{with isospin}~~ I=3/2$,}

\vspace{0.2cm}
\noindent $S=-2,~ \text{channels}: \quad  \bar{D}_s \Lambda,~ \bar{D} \Xi,\quad I=0$;\quad \R{ $\bar{D} \Xi,\quad ~~ I=1$,}

\vspace{0.2cm}
\noindent $S=-3,~ \text{channel}: \quad  \bar{D}_s \Xi,\quad I=1/2$,

\vspace{0.2cm}
\R{
\begin{itemize}
\item \textbf{Decuplet baryon}s:
\end{itemize}
}
{\color{black}
\noindent $S=-1,~ \text{channels}: \quad  \bar{D} \Sigma^*,\quad \text{with isospin}~~ I=1/2$; \quad $\bar{D}_s \Delta,~  \bar{D} \Sigma^*,\quad \text{with isospin}~~ I=3/2$;

\vspace{0.2cm}
\noindent $S=-2,~ \text{channels}: \quad  \bar{D} \Xi^*,\quad ~~ I=0\quad  \bar{D}_s \Sigma^*,~ \bar{D} \Xi^*,\quad I=1$;

\vspace{0.2cm}
\noindent $S=-3,~ \text{channel}: \quad  \bar{D}_s \Xi^*,~\bar{D} \Omega\quad I=1/2$,
}

\vspace{0.2cm}
\noindent $S=-4,~ \text{channel}: \quad  \bar{D}_s \Omega,\quad I=0$.\\

\noindent We construct states with $I=1/2$ and $I=0$ using isospin multiplets:
\begin{align}
&D=\binom{D^{+}}{-D^0},\quad \bar{D}=\binom{\bar D^{0}}{D^-},\quad K=\binom{K^{+}}{K^0},\quad\bar{K}=\binom{\bar{K}^0}{-K^{-}},\quad \Xi=\binom{\Xi^0}{-\Xi^{-}},\quad  \Sigma = \begin{pmatrix}
-\Sigma^+ \\
\Sigma^0 \\
\Sigma^-
\end{pmatrix},\\
&\R{\Xi=\binom{\Xi^{*0}}{\Xi^{*-}},\quad  \Sigma = \begin{pmatrix}
\Sigma^{*+} \\
\Sigma^{*0} \\
\Sigma^{*-}
\end{pmatrix}}.\\
\end{align}
This leads to the following isospin states:

\begin{eqnarray}
| \bar D \Sigma , ~I=1/2, ~I_3=1/2 \rangle &=& \frac{1}{\sqrt{3}}|\bar D^0\Sigma^0 \rangle + \frac{\sqrt{2}}{\sqrt{3}}| D^-\Sigma^+ \rangle, \nonumber\\
| \bar D \Xi , ~I=0 \rangle &=& -\frac{1}{\sqrt{2}}\left(|\bar D^0\Xi^- \rangle + |D^-\Xi^0 \rangle\right), \nonumber\\
{\color{black} | \bar D \Sigma^* , ~I=1/2, ~I_3=1/2 \rangle } &=& {\color{black} \frac{1}{\sqrt{3}}|\bar D^0\Sigma^{*0} \rangle - \frac{\sqrt{2}}{\sqrt{3}}| D^-\Sigma^{*+} \rangle,} \nonumber\\
{\color{black} | \bar D \Xi^* , ~I=0 \rangle } &=& {\color{black} \frac{1}{\sqrt{2}}\left(|\bar D^0\Xi^{*-} \rangle - |D^-\Xi^{*0} \rangle\right)}. \nonumber
\end{eqnarray}

We shall use the local hidden gauge approach~\cite{Bando:1984ej,Bando:1987br,Meissner:1987ge,Nagahiro:2008cv}, exchanging vector mesons, in order to evaluate the interaction of these systems.
With the former wave functions and the prescriptions to calculate the $VPP$ and $VBB$ vertices ($V\equiv$ vector, $P\equiv$ pseudoscalar, $B\equiv$ baryon), we can build the transition potential matrix elements for the channels listed above. Details on the calculations can be seen, for instance, in the appendix of Ref.~\cite{Debastiani:2017ewu}.

\subsection{Interaction between Coupled Channels}
\begin{figure}[H]
  \centering
  \includegraphics[width=4.5cm]{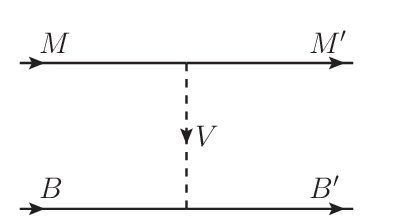}
  \caption{Diagrammatic representation of the interaction $MB\to M^\prime B^\prime$ via vector meson exchange. Here, $M$ ($M^\prime$) and $B$ ($B^\prime$) are the initial (final) mesons and baryons, respectively, and $V$ is the exchanged vector meson.}
  \label{fig1}
\end{figure}

In the hidden gauge approach, the meson-baryon interaction within SU(3) is obtained through the exchange of vector mesons, as illustrated in Fig.~\ref{fig1},

The meson-baryon interaction is described using vector meson exchange in the framework of the extended local hidden gauge approach. The $VPP$ vertex is given by the Lagrangian:
\begin{align}
\mathcal{L}_{\mathrm{VPP}} = -i g\left\langle\left[P, \partial_{\mu} P\right] V^{\mu}\right\rangle, \label{lPPV}
\end{align}
with $g = \frac{m_V}{2f_\pi}$, where $m_V = 800$ MeV and $f_\pi = 93$ MeV.
Here, $P$ and $V$ are the SU(4) pseudoscalar and vector meson matrices, using the $\eta,~\eta'$  mixing of Ref.~\cite{Bramon:1992kr}:
\begin{equation}\label{PMat}
P =
\begin{pmatrix}
\frac{1}{\sqrt{2}} \pi^0 + \frac{1}{\sqrt{3}} \eta + \frac{1}{\sqrt{6}} \eta' & \pi^+ & K^+ & \bar{D}^0 \\
\pi^- & -\frac{1}{\sqrt{2}} \pi^0 + \frac{1}{\sqrt{3}} \eta + \frac{1}{\sqrt{6}} \eta' & K^0 & D^- \\
K^- & \bar{K}^0 & -\frac{1}{\sqrt{3}} \eta + \sqrt{\frac{2}{3}} \eta' & D_s^- \\
D^0 & D^+ & D_s^+ & \eta_c
\end{pmatrix},
\end{equation}
\begin{equation}\label{VMat}
V =
\begin{pmatrix}
\frac{1}{\sqrt{2}} \rho^0 + \frac{1}{\sqrt{2}} \omega & \rho^+ & K^{*+} & \bar{D}^{*0} \\
\rho^- & -\frac{1}{\sqrt{2}} \rho^0 + \frac{1}{\sqrt{2}} \omega & K^{*0} & \bar{D}^{*-} \\
K^{*-} & \bar{K}^{*0} & \phi & D_s^{*-} \\
D^{*0} & D^{*+} & D_s^{*+} & J/\psi
\end{pmatrix}.
\end{equation}

The vector-baryon coupling for baryons of the SU(3) octet is described by the Lagrangian:
\begin{equation}
\mathcal{L}_{BBV'} = g\left( \langle \bar{B}\gamma_{\mu}[V'^{\mu},B] \rangle + \langle \bar{B}\gamma_{\mu}B \rangle \langle V'^{\mu} \rangle \right),
\label{lbbv}
\end{equation}

where $V'$, $B$ are the $\textrm{SU(3)}$ matrices for vector mesons and baryons
\begin{equation}
\label{vfields}
V'=\left(
\begin{array}{ccc}
\frac{1}{\sqrt{2}}\rho^0+\frac{1}{\sqrt{2}}\omega & \rho^+ & K^{*+}\\
\rho^- &-\frac{1}{\sqrt{2}}\rho^0+\frac{1}{\sqrt{2}}\omega & K^{*0}\\
K^{*-} & \bar{K}^{*0} &\phi\\
\end{array}
\right)\ ,
\end{equation}
\begin{equation}
B =
\begin{pmatrix}
\frac{1}{\sqrt{2}} \Sigma^0 + \frac{1}{\sqrt{6}} \Lambda & \Sigma^+ & p \\
\Sigma^- & -\frac{1}{\sqrt{2}} \Sigma^0 + \frac{1}{\sqrt{6}} \Lambda & n \\
\Xi^- & \Xi^0 & -\frac{2}{\sqrt{6}} \Lambda
\end{pmatrix}.
\end{equation}
At low energies, the dominant contributions come from the time components $\partial_0$ and $\gamma^0$~\cite{Oller:2000ma}. The resulting interaction kernel reads:
\begin{equation}
V_{ij} = C_{ij} \frac{1}{4f_\pi^2}(k^0 + k^{\prime 0}),
\label{kernel}
\end{equation}
where $k^0,~k^{\prime 0}$  are the meson energies of the initial, final state, given by,
\begin{equation}
k^0 = \frac{s + m_{m_i}^2 - M_{B_i}^2}{2\sqrt{s}}, \quad k^{\prime 0} = \frac{s + m_{m_j}^2 - M_{B_j}^2}{2\sqrt{s}}.
\end{equation}
Here, $m_{m_i}, M_{B_i}$ and $m_{m_j}, M_{B_j}$ denote the meson and baryon masses in the initial and final states, respectively. The coefficients $C_{ij}$ are easily obtained from the vertices of  Eqs.~(\ref{lPPV}) and ~(\ref{lbbv}) with the Feynman diagram of Fig.~\ref{fig1}. 

In our analysis, we consider meson-baryon interactions in various strangeness sectors, denoted by the quantum number \( S \). The coupled channels involved and their corresponding interaction strengths are characterized by the coefficients \( C_{ij} \). The explicit values of \( C_{ij} \) \R{coefficients involving channels with octet baryons in each strangeness sector are summarized in} Tables~\ref{coeff_S1},~\ref{coeff_S2},~\ref{coeff_S3}, and~\ref{coeff_S3_1}.

In $S=-1$ sector
\begin{table}[H]
\centering
 \caption{Coefficients $C_{ij}$ for the $S=-1$ sector with isospin $I=1/2$.}
 \label{coeff_S1}
\setlength{\tabcolsep}{6.5pt}
\begin{tabular}{l|ccc}
\hline
\hline
          ~           & ~$\bar{D}_s N$~ & ~$\bar{D} \Lambda$~ & ~$\bar{D} \Sigma$\\
\hline
$\bar{D}_s N$~       &        $0$         & $-\sqrt{\frac{3}{2}}$  & $-\sqrt{\frac{3}{2}}$  \\
$\bar{D} \Lambda$~   &                 &            $1$         &          $0$        \\
$\bar{D} \Sigma$~    &                 &                     &           $-1$       \\
\hline
\hline
\end{tabular}
\end{table}
{\color{black}
 \begin{equation}
C(\bar{D} \Sigma, I = 3/2) = 2.
\end{equation} }

In $S=-2$ sector
\begin{table}[H]
\centering
 \caption{Coefficients $C_{ij}$ for the $S=-2$ sector.}
 \label{coeff_S2}
\setlength{\tabcolsep}{6.5pt}
\begin{tabular}{l|cc}
\hline
\hline
          ~                 & ~$\bar{D}_s \Lambda$~ & ~$\bar{D} \Xi$\\
\hline
$\bar{D}_s \Lambda$~       &     $1$                 &       $-\sqrt{3} $      \\
$\bar{D} \Xi$~             &                      &        $-1$      \\
\hline
\hline
\end{tabular}
\end{table}
{\color{black}
 \begin{equation}
C(\bar{D}\Xi, I = 1) = 1.
\end{equation} }

In $S=-3$ sector
\begin{table}[H]
\centering
 \caption{Coefficient $C_{ij}$ for the $S=-3$ sector.}
 \label{coeff_S3}
\setlength{\tabcolsep}{6.5pt}
\begin{tabular}{l|c}
\hline
\hline
          ~           & ~$\bar{D}_s \Xi$ \\
\hline
$\bar{D}_s \Xi$ ~     &            $2 $     \\
\hline
\hline
\end{tabular}
\end{table}
The evaluation of the lower vertex for vector couplings to decuplet baryons is done using explicit quark wave functions  of the vectors and the baryons as shown in the Appendix.
{\color{black}
Then the coefficients corresponding to the channels with decuplet baryons are given in Tables~\ref{coeff_decuplet_S1}, ~\ref{coeff_decuplet_S2} and~\ref{coeff_decuplet_S3}. 

In $S=-1$ sector 
\begin{equation}
C(\bar{D} \Sigma^*, I = \tfrac{1}{2}) = -1.
\end{equation} }
\begin{table}[H]
\centering
 \caption{Coefficients $C_{ij}$ for the $S=-1$ sector with isospin $I=3/2$.}
 \label{coeff_decuplet_S1}
\setlength{\tabcolsep}{6.5pt}
{\color{black}
\begin{tabular}{l|cc}
\hline
\hline
          ~           & ~$\bar{D}_s \Delta$~  & ~$\bar{D} \Sigma^*$\\
\hline
$\bar{D}_s \Delta$~       &        $0$         & $\sqrt{{3}}$    \\
$\bar{D} \Sigma^*$~    &                 &                                $2$       \\
\hline
\hline
\end{tabular}}
\end{table}
{\color{black}
In $S=-2$ sector \begin{equation}
C(\bar{D} \Xi^*, I = 0) = -1.
\end{equation} }
\begin{table}[H]
\centering
 \caption{Coefficients $C_{ij}$ for the $S=-2$ sector with isospin $I=1$.}
 \label{coeff_decuplet_S2}
\setlength{\tabcolsep}{6.5pt}
{\color{black}
\begin{tabular}{l|cc}
\hline
\hline
          ~                 & ~$\bar{D}_s \Sigma^*$~ & ~$\bar{D} \Xi^*$\\
\hline
$\bar{D}_s \Lambda$~       &     $1$                 &       $2 $      \\
$\bar{D} \Xi$~             &                      &        $1$      \\
\hline
\hline
\end{tabular}
}
\end{table}

In $S=-3$ sector 
\begin{table}[H]
\centering
 \caption{Coefficients $C_{ij}$ for the $S=-3$ sector with isospin $I=1/2$.}
 \label{coeff_decuplet_S3}
\setlength{\tabcolsep}{6.5pt}
{\color{black}
\begin{tabular}{l|cc}
\hline
\hline
          ~                 & ~$\bar{D}_s \Xi^*$~ & ~$\bar{D} \Omega$\\
\hline
$\bar{D}_s \Lambda$~       &     $2$                 &       $\sqrt{3} $      \\
$\bar{D} \Xi$~             &                      &        $0$      \\
\hline
\hline
\end{tabular}
}
\end{table}
In $S=-4$ sector 
\begin{table}[H]
\centering
 \caption{Coefficient $C_{ij}$ for the $S=-4$ sector.}
 \label{coeff_S3_1}
\setlength{\tabcolsep}{6.5pt}
\begin{tabular}{l|c}
\hline
\hline
          ~           & ~$\bar{D}_s \Omega$ \\
\hline
$\bar{D}_s \Omega$ ~  &        $3 $           \\
\hline
\hline
\end{tabular}
\end{table}

By using Eq.~(\ref{lPPV}) with the $q \bar q$ matrices written in term of pseudoscalar mesons, $P$ in Eq.~(\ref{PMat}), or vector mesons, $V$ in  Eq.~(\ref{VMat}), it looks like one is using SU(4) symmetry. However, this is not the case because, since the baryons have no charm, we cannot exchange any vector meson containing  charm. Thus, the $\bar c$ quark of the mesons acts as a spectator and one only exchanges $\rho,~\omega,~K^*$ mesons. In summary, one is making use of the SU(3) substructure of the Lagrangian that we use. The lower vertex of Eq.~(\ref{lbbv})  uses SU(3) symmetry directly.

\subsection{Scattering Matrix and Pole Structure}
\label{BSE}

Once the interaction potential $V_{ij}$ is obtained, the scattering amplitude is calculated by solving the coupled-channel Bethe-Salpeter equation:
\begin{equation}
T = [1 - V G]^{-1} V,
\end{equation}
where $G$ is the diagonal matrix of meson-baryon loop functions, given by:
\begin{equation}
G_i (\sqrt{s}) = 2M_i \int_{|\vec{q}| < q_{\text{max}}} \frac{d^3 q}{(2\pi)^3} \frac{\omega_1(\vec{q}) + \omega_2(\vec{q})}{2 \omega_1(\vec{q}) \omega_2(\vec{q})} \frac{1}{s - [\omega_1(\vec{q}) + \omega_2(\vec{q})]^2 + i\epsilon},
\label{eq:G}
\end{equation}
with
\begin{equation}
\omega_1(\vec{q}) = \sqrt{\vec{q}^{\,2} + m_i^2}, \quad \omega_2(\vec{q}) = \sqrt{\vec{q}^{\,2} + M_i^2}.
\end{equation}
We adopt a cutoff regularization with $q_{\text{max}} = 630$ MeV, consistent with Ref.~\cite{Oset:1997it}, but vary $q_{\text{max}}$ is a reasonable range to get uncertainties also taking $q_{\text{max}} = 600$ MeV from the study of the $P_{cs}$ states~\cite{Feijoo:2022rxf} and $q_{\text{max}} = 650$ MeV from the study of $\Omega_c$ states~\cite{Debastiani:2017ewu}.

To search for resonances, we continue the amplitude analytically to the second Riemann sheet by replacing $G \rightarrow G^{II}$ for open channels (see Ref.~\cite{Debastiani:2017ewu}), but since the bound states  obtained are below all thresholds, the poles appear in the first Riemann sheet.
The poles are reached in the second Riemann sheet for which we change $G\to G^{II}$ as
\begin{eqnarray}
G^{II}_j = G^I_j + i \frac{2M_j\,q}{4\pi\sqrt{s}}\,,
\end{eqnarray}
for Re$\sqrt s>m_j+M_j$, and $q$ is
\begin{eqnarray}
q = \frac{\lambda^{1/2}(s,m^2_j,M^2_j)}{2\sqrt{s}}\,,
\end{eqnarray}
with $m_j$ and $M_j$ the masses of the meson and baryon, respectively. We also evaluate the couplings defined from the residues 
at the pole, where the amplitudes go as 
\begin{equation}
T_{ij} = \frac{g_i g_j}{z-z_R} \, ,
\end{equation}
with $z_R$ the complex energy ($M, i\Gamma/2$). We choose one sign for one $g_1$ and the rest of the couplings have the 
relative sign well defined~\cite{Gamermann:2009uq}.

\section{Results} 
\label{res}

We first summarize the threshold masses of the meson-baryon channels considered in this work, as listed in Tables~\ref{tab1}, \ref{tab2}, \ref{tab3}, and \ref{tab3_1}, the masses of mesons and baryons are from PDG~\cite{ParticleDataGroup:2024cfk}. These thresholds provide the baseline for understanding possible bound or resonance states generated by the interactions.
\begin{table}[H]
\caption{Threshold masses (in MeV) of different channels for $S=-1$.}
\centering
\begin{tabular}{c | c c c c c}
\hline\hline
{\bf States} & $\bar{D}_s N$ & $\bar{D} \Lambda$ & $\bar{D} \Sigma$ & ~$\bar{D}_s \Delta$~  & ~$\bar{D} \Sigma^*$\\
\hline
{\bf Threshold} & $2906.62$ & $2980.52$ & $3060.4$ & $3200.35$ & $3250.08$ \\
\hline\hline
\end{tabular}
\label{tab1}
\end{table}

\begin{table}[H]
\caption{Threshold masses (in MeV) of different channels for $S=-2$.}
\centering
\begin{tabular}{c | c c cc}
\hline\hline
{\bf States} & $\bar{D}_s \Lambda$ & $\bar{D} \Xi$ & ~$\bar{D}_s \Sigma^*$~ & ~$\bar{D} \Xi^*$\\
\hline
{\bf Threshold} & $3084.03$ & $3185.54$ & $3351.18$ & $3399.05$ \\
\hline\hline
\end{tabular}
\label{tab2}
\end{table}

\begin{table}[H]
\caption{Threshold mass (in MeV) of different channel for $S=-3$.}
\centering
\begin{tabular}{c | ccc }
\hline\hline
{\bf States} & $\bar{D}_s \Xi$ & ~$\bar{D}_s \Xi^*$~ & ~$\bar{D} \Omega$\\
\hline
{\bf Threshold} & $3283.21$ & $3500.15$ & $3537.29$ \\
\hline\hline
\end{tabular}
\label{tab3}
\end{table}

\begin{table}[H]
\caption{Threshold mass (in MeV) of different channel for $S=-4$.}
\centering
\begin{tabular}{c | c }
\hline\hline
{\bf States} & $\bar{D}_s \Omega$ \\
\hline
{\bf Threshold} & $3640.8 $\\
\hline\hline
\end{tabular}
\label{tab3_1}
\end{table}
Using the formalism described above, we solve the Bethe-Salpeter equation with the interaction kernels constructed from the vector meson exchange model. The interaction proceeds via $S-$wave. The coefficients $C_{ij}$ for each strangeness sector ($S=-1, -2, -3, -4$) determine the strength and nature of the coupled-channel interactions.

\subsection{Results for the $S=-1$ sector}
\R{
\begin{itemize}
\item \textbf{Octet baryon}s:
\end{itemize}
}

\R{In the \( S = -1 \) sector, we consider the coupled channels \(\bar{D}_s N\), \(\bar{D} \Lambda\), and \(\bar{D} \Sigma\) with isospin \( I = 1/2 \). The interactions among these channels are attractive as a consequence of the attraction in the  \(\bar{D} \Sigma\)  channel and the role of coupled channels, resulting in a dynamically generated bound state in the scattering amplitude. It should be noted that no states with isospin \( I = 3/2 \) appear in this sector for \(\bar{D} \Sigma\) in $I = 3/2$ since the interaction is repulsive. To estimate the uncertainties in our results, several different values of the cutoff parameter \( q_{\max} \) have been used.
}

The coupling constants of the pole to each channel are listed in Table~\ref{S1POLE}, together with the wave function at the origin~\cite{Gamermann:2009uq},  $g_iG_i$. The strongest coupling is to the $\bar{D} \Sigma$ channel, followed by $\bar{D}_s N$ and $\bar{D} \Lambda$. This suggests that the dominant component of the bound state is $\bar{D} \Sigma$.
However, as found in~\cite{Gamermann:2009uq}, the wave function at the origin, which {enters} the evaluation at processes involving the short range strong interaction,  is
\begin{align}
    \Psi_i(0)\sim g_iG_i\bigg|_\text{pole},
\end{align}
indicating that the most relevant channel is $\bar{D}_s N$, but the other channels, particularly the $\bar{D} \Sigma$  are not negligible.

\begin{table}[H]
\centering
\caption{The pole (in MeV) and its coupling constants $g_i$ and $g_iG_i$  (in MeV) to the channels in the $S=-1$ sector with issopin $I=1/2$.}
\label{S1POLE}
\setlength{\tabcolsep}{28pt}
\begin{tabular}{ll|lccc}
\hline \hline
    $q_\text{max}$ &     Pole          &       ~            & $\bar{D}_s N$ & $\bar{D} \Lambda$ & $\bar{D} \Sigma$ \\
\hline
$550$  & $\textbf{2906 }$  & $g_i$& $0.91 $  & $0.72 $  & $1.43 $  \\
   &              & $g_iG_i$ & $-7.62 $ & $-2.85 $  & $-3.94 $     \\
\cline{3-6}
$600$  & $\textbf{2898 }$  & $g_i$& $2.51 $  & $1.64 $  & $3.59 $  \\
   &              & $g_iG_i$ & $-17.89 $ & $-7.23 $  & $-11.52 $     \\
\cline{3-6}
$630$  &     \textbf{2888}       & $g_i$              & $3.10$        & $1.85$            & $4.26$           \\
             &   & $g_iG_i$              & $-20.43$              & $-8.45$    & $-14.71$     \\
\cline{3-6}
$650$  & $\textbf{2880  }$  & $g_i$& $3.40 $  & $1.95 $  & $4.59 $  \\
   &              & $g_iG_i$ & $-21.7 $ & $-9.09 $  & $-16.39 $     \\
\hline
\hline
\end{tabular}
\end{table}
\R{
\begin{itemize}
\item \textbf{Decuplet baryon}s:
\end{itemize}
}
\R{
In the sector involving decuplet baryons, the channel \(\bar{D} \Sigma^*\) with isospin \( I = \frac{1}{2} \) generates a pole in the scattering amplitude. To estimate the uncertainties of this result, several different values of the cutoff parameter \( q_{\max} \) have been considered. The results are summarized in Table~\ref{NEW_1}.}

\begin{table}[H]
\centering
\caption{The pole (in MeV) and its coupling constants $g_i$ and $g_iG_i$  (in MeV)  to the channels in the $I=1/2,~S=-1$ sector.}
\label{NEW_1}
\setlength{\tabcolsep}{28pt}
\begin{tabular}{ll|lcc}
\hline \hline
    $q_\text{max}$ &     Pole          &       ~            &  $\bar{D} \Sigma^*$   \\
\hline
{550 } & $\textbf{3246.38 }$  & $g_i$&  $1.86 $   \\
   &              & $g_iG_i$ & $-17.25  $       \\
\cline{3-4}
{600 } & $\textbf{3241.99 }$  & $g_i$&  $2.35 $   \\
   &              & $g_iG_i$ & $-21.82  $       \\
\cline{3-4}
{630 } & $\textbf{3238.39 }$  & $g_i$&  $2.63 $   \\
   &              & $g_iG_i$ & $-24.47  $ \\ \cline{3-4}
{650 } & $\textbf{3235.56 }$  & $g_i$&  $2.82 $   \\
   &              & $g_iG_i$ & $-26.22  $       \\
\hline
\hline
\end{tabular}
\end{table}

{\color{black}
Tables~\ref{S1POLE} and \ref{NEW_1} summarize the pole positions (in MeV) and their coupling constants \( g_i \) and \( g_i G_i \) (in MeV) to the relevant channels in the \( S = -1 \) sector with isospin \( I = \frac{1}{2} \).
Table~\ref{S1POLE} corresponds to the octet baryon channels \(\bar{D}_s N\), \(\bar{D} \Lambda\), and \(\bar{D} \Sigma\). As the cutoff \( q_{\max} \) increases from 550 MeV to 650 MeV, the pole position shifts from 2906 MeV to 2880 MeV, showing a stronger binding. The couplings indicate that the \(\bar{D} \Sigma\) channel dominates the bound state formation.
Table~\ref{NEW_1} displays the results for the decuplet baryon channel \(\bar{D} \Sigma^*\), with a threshold at 3250.08 MeV. Varying \( q_{\max} \) between 550 MeV and 650 MeV, the pole moves from 3246.38 MeV down to 3235.56 MeV, indicating increasing binding strength. The coupling constants also grow in magnitude accordingly, consistent with the enhanced interaction strength.

We see in Table~\ref{S1POLE} that we always get a bound state below the  \(\bar{D}_s N\) threshold and the uncertainties in the binding are about 26 MeV. We also see in Table~\ref{NEW_1} that bound states are generated in all cases and the uncertainty in the binding is of about 10 MeV.

Together, these results demonstrate the presence of dynamically generated states in both octet and decuplet baryon sectors, with their properties sensitive to the cutoff parameter used in the calculation.

For the isospin \( I = 3/2 \) channel, as shown by the coefficients in Table~\ref{coeff_decuplet_S1}, the interaction potential is repulsive. However, when the cutoff parameter \( q_{\max} \) exceeds 630 MeV, a weakly bound state appears near the lowest threshold as a consequence of the attraction produced by the coupled channels (see section~\ref{dis} for more details on the effect of coupled channels).  As \( q_{\max} \) decreases, this structure disappears because the effective potential becomes positive, indicating a repulsive interaction.

}

\subsection{Results for the $S=-2$ sector}

\R{
In the \( S = -2 \) sector, we consider the coupled channels \(\bar{D}_s \Lambda\) and \(\bar{D} \Xi\) with isospin \( I = 0 \). The interaction in this channel is attractive, leading to the formation of a bound state below threshold. To estimate the uncertainties in our results, several different values of the cutoff parameter \( q_{\max} \) have been employed. It should be noted that no bound states or resonant structures appear in the isospin \( I = 1 \) channel of \(\bar{D} \Xi\).
The detailed results for the \( S = -2 \) sector with isospin \( I=0 \) are presented in Table~\ref{S2POLE}.
}


\begin{table}[H]
\centering
\caption{The pole (in MeV) and its coupling constants $g_i$ and $g_iG_i$  (in MeV)  to the channels in the $S=-2$ sector with isospin $I=0$.}
\label{S2POLE}
\setlength{\tabcolsep}{28pt}
\begin{tabular}{ll|lcc}
\hline \hline
    $q_\text{max}$ &     Pole          &       ~            & $\bar{D}_s \Lambda$ & $\bar{D} \Xi$  \\
\hline
{560 } & $\textbf{3083 }$  & $g_i$& $0.92 $  & $2.71 $   \\
   &              & $g_iG_i$ & $-8.70 $ & $-10.05 $       \\
\cline{3-5}
$600$  & $\textbf{3071  }$  & $g_i$& $2.12 $  & $5.15 $  \\
   &              & $g_iG_i$ & $-15.72 $ & $-20.63 $       \\
\cline{3-5}
$630$  &  \textbf{3057}       & $g_i$              & $2.55$              & $5.72$         \\
  &    & $g_iG_i$              & $-17.09$              & $-23.73$         \\   
\cline{3-5}
$650$  & $\textbf{3046  }$  & $g_i$& $2.79$  & $ 6.03$    \\
   &              & $g_iG_i$ & $-17.82 $ & $-25.49 $       \\
\hline
\hline
\end{tabular}
\end{table}
The bound state couples more strongly to the $\bar{D} \Xi$ channel  than to the $\bar{D}_s \Lambda$ channel, indicating that the $\bar{D} \Xi$ component dominates the internal structure of this state, but the wave functions at the origin are more similar.

\R{It should be noted that the smallest cutoff value \( q_{\max} \) considered in our analysis is 560 MeV, where a barely bound state with respect to the $\bar{D}_s \Lambda$   threshold appears. If we decrease \(q_{\text{max}}\) to values below 550 MeV, the pole moves above this threshold becoming a resonance. For $\bar{D} \Xi$  in $I=1$ the interaction is repulsive and we do not get any bound state.
}
\R{
\begin{itemize}
\item \textbf{Decuplet baryon}s:
\end{itemize}
}
\R{
In the sector involving decuplet baryons, the channel \(\bar{D} \Xi^*\) with isospin \( I = 0 \) generates a pole in the scattering amplitude. To estimate the uncertainties of this result, several different values of the cutoff parameter \( q_{\max} \) have been considered. The results are summarized in Table~\ref{NEW_2}.}

\begin{table}[H]
\centering
\caption{The pole (in MeV) and its coupling constants $g_i$ and $g_iG_i$  (in MeV)  to the channels in the $I=0,~S=-2$ sector. }
\label{NEW_2}
\setlength{\tabcolsep}{28pt}
\begin{tabular}{ll|lcc}
\hline \hline
    $q_\text{max}$ &     Pole          &       ~            &  $\bar{D} \Xi^*$    \\
\hline
{550 } & $\textbf{3393  }$  & $g_i$&  $2.03 $   \\
   &              & $g_iG_i$ & $-18.80  $       \\
\cline{3-4}
{600 } & $\textbf{3388 }$  & $g_i$&  $2.49 $   \\
   &              & $g_iG_i$ & $-23.16  $       \\
\cline{3-4}
{630 } & $\textbf{3383  }$  & $g_i$&  $2.77 $   \\
   &              & $g_iG_i$ & $-25.72  $       \\
\cline{3-4}
{650 } & $\textbf{3380 }$  & $g_i$&  $2.95 $   \\
   &              & $g_iG_i$ & $-27.42  $       \\
\hline
\hline
\end{tabular}
\end{table}
{\color{black}
Table~\ref{S2POLE} shows the results for the octet baryon channels \(\bar{D}_s \Lambda\) and \(\bar{D} \Xi\). We get bound states in this case. We see that now the binding energies change more than in the former cases with the cutoff parameter \(q_{\max}\), reflecting the sensitivity of the bound state properties to the model parameters.
Table~\ref{NEW_2} corresponds to the decuplet baryon channel \(\bar{D} \Xi^*\) with $I=0$. Now the changes in \(q_{\max}\) only induce a change of 13 MeV in the binding energy.

}
\subsection{Results for the $S=-3$ and $S=-4$ sector}

\R{In the \(S=-3\) sector, we consider both octet and decuplet baryon systems. For the octet case, there is a single channel, \(\bar{D}_s \Xi\). In the decuplet case with isospin \(I=1/2\), the coupled channels \(\bar{D}_s \Xi^*\) and \(\bar{D} \Omega\) are taken into account. These configurations allow us to investigate possible dynamically generated states resulting from the meson–baryon interactions in this sector.} 
The interaction is found to be repulsive. The $C_{ij}$ coefficient are positive, indicating a repulsive force which cannot bind the system. The same occurs in the  $S=-4$ sector with the $\bar{D}_s \Omega$ channel. As a result, we do not find any poles in the scattering amplitudes below threshold nor in the complex plane near the real axis. This confirms the absence of bound or resonant states in the $S=-3,-4$ sectors within the framework of our model.

\subsection{Vector baryon interaction}
It is easy to obtain similar results for systems where one replaces $\bar D,~\bar D_s$ by $\bar D^*,~\bar D^*_s$. The transition potentials are identical to those used alone, except for the change of the masses~\cite{Wang:2022aga}. Then we show in Tables~\ref{tab1_stc},~\ref{tab2_stc},~\ref{tab3_stc}, and~\ref{tab3_1_stc} the thresholds of the coupled channels involving these vector mesons,
\begin{table}[H]
\caption{Threshold masses (in MeV) of different channels for $S=-1$.}
\centering
\begin{tabular}{c | c c c cc }
\hline\hline
{\bf States} & $\bar{D}^*_s N$ & $\bar{D}^* \Lambda$ & $\bar{D}^* \Sigma$ & ~$\bar{D}^*_s \Delta$~  & ~$\bar{D}^* \Sigma^*$  \\
\hline
{\bf Threshold} & $3050.47$ & $3122.53$ & $3.20171 $ & $3344.2 $ & $3391.39 $ \\
\hline\hline
\end{tabular}
\label{tab1_stc}
\end{table}
\begin{table}[H]
\caption{Threshold masses (in MeV) of different channels for $S=-2$.}
\centering
\begin{tabular}{c | c c cc}
\hline\hline
{\bf States} & $\bar{D}^*_s \Lambda$ & $\bar{D}^* \Xi$ & ~$\bar{D}^*_s \Sigma^*$~ & ~$\bar{D}^* \Xi^*$ \\
\hline
{\bf Threshold} & $3227.88$ & $3326.84$ & $3495.03 $ & $3540.36 $ \\
\hline\hline
\end{tabular}
\label{tab2_stc}
\end{table}

\begin{table}[H]
\caption{Threshold mass (in MeV) of different channel for $S=-3$.}
\centering
\begin{tabular}{c | ccc }
\hline\hline
{\bf States} & $\bar{D}^*_s \Xi$ & ~$\bar{D}_s \Xi^*$~ & ~$\bar{D} \Omega$ \\
\hline
{\bf Threshold} & $3430.49$ & $3644 $ & $ 3681.01$ \\
\hline\hline
\end{tabular}
\label{tab3_stc}
\end{table}

\begin{table}[H]
\caption{Threshold mass (in MeV) of different channel for $S=-4$.}
\centering
\begin{tabular}{c | c }
\hline\hline
{\bf States} & $\bar{D}^*_s \Omega$ \\
\hline
{\bf Threshold} & $3784.65$ \\
\hline\hline
\end{tabular}
\label{tab3_1_stc}
\end{table}

\R{
\begin{itemize}
\item \textbf{Octet baryon}s:
\end{itemize}
}
\R{
In Tables~\ref{S1POLE_stc} and~\ref{S2POLE_stc}, we show the poles, couplings, and wave functions at the origin for the states that we obtain, taking into account different values of the cutoff parameter \(q_{\text{max}}\) to estimate the associated uncertainties.
}

\begin{table}[H]
\centering
\caption{The pole (in MeV) and its coupling constants $g_i$ and $g_iG_i$  (in MeV) to the channels in the $S=-1$ sector.}
\label{S1POLE_stc}
\setlength{\tabcolsep}{28pt}
\begin{tabular}{ll|lccc}
\hline \hline
    $q_\text{max}$ &     Pole          &       ~            & $\bar{D}^*_s N$ & $\bar{D}^* \Lambda$ & $\bar{D}^* \Sigma$ \\
\hline
$550$  & $\textbf{3049  }$  & $g_i$& $1.37 $  & $1.06 $  & $2.13 $  \\
   &              & $g_iG_i$ & $-10.44 $ & $-3.96 $  & $-5.55 $     \\
\cline{3-6}
$600$  & $\textbf{3039 }$  & $g_i$& $2.74 $  & $1.75 $  & $3.89 $  \\
   &              & $g_iG_i$ & $-17.87 $ & $-7.27 $  & $-11.76 $     \\
\cline{3-6}
$630$  &      \textbf{3028 }       & $g_i$              & $3.31$        & $1.94$            & $4.53$           \\
            &    & $g_iG_i$              & $-20.05$              & $-8.33$    & $-14.71$     \\
\cline{3-6}
$650$  & $\textbf{3020}$  & $g_i$& $3.62 $  & $ 2.04$  & $4.86 $  \\
   &              & $g_iG_i$ & $-21.19 $ & $-8.91 $  & $-16.28 $     \\   
\hline
\hline
\end{tabular}
\end{table}

\begin{table}[H]
\centering
\caption{The pole (in MeV) and its coupling constants $g_i$ and $g_iG_i$  (in MeV)  to the channels in the $S=-2$ sector.}
\label{S2POLE_stc}
\setlength{\tabcolsep}{28pt}
\begin{tabular}{ll|lcc}
\hline \hline
    $q_\text{max}$ &     Pole          &       ~            & $\bar{D}^*_s \Lambda$ & $\bar{D}^* \Xi$  \\
\hline
$550$  & $\textbf{3227  }$  & $g_i$& $0.86 $  & $2.59 $    \\
   &              & $g_iG_i$ & $-7.73 $ & $-8.84 $       \\
\cline{3-5}
\cline{3-5}
$600$  & $\textbf{3211 }$  & $g_i$& $2.31 $  & $5.48 $    \\
   &              & $g_iG_i$ & $-15.42 $ & $-20.57 $      \\
\cline{3-5}
$630$  & \textbf{3196}       & $g_i$              & $2.72$              & $6.01$         \\
    &  & $g_iG_i$              & $-16.59$              & $-23.33$         \\
\cline{3-5}
$650$  & $\textbf{3184 }$  & $g_i$& $2.97 $  & $6.32 $   \\
   &              & $g_iG_i$ & $-17.27 $ & $-24.96 $      \\
\hline
\hline
\end{tabular}
\end{table}
{\color{black}
Tables~\ref{S1POLE_stc} and~\ref{S2POLE_stc} show the poles, coupling constants, and wave functions at the origin for the states found in the vector-baryon interaction sectors with strangeness $S=-1$ and $S=-2$, respectively. In the $S=-1$ sector, the coupled channels $\bar{D}^*_s N$, $\bar{D}^* \Lambda$, and $\bar{D}^* \Sigma$ dynamically generate a bound state. Among these, the state couples most strongly to the $\bar{D}^*_s N$ channel, as indicated by the largest magnitude of $g_i G_i$, suggesting that this component dominates in the wave function at the origin.

In the $S=-2$ sector, the coupled-channel dynamics of $\bar{D}^*_s \Lambda$ and $\bar{D}^* \Xi$ also lead to a bound state, with the $\bar{D}^* \Xi$ channel playing the dominant role. From the results, we observe that the vector–baryon interactions tend to produce more deeply bound states than their pseudoscalar–baryon counterparts, a trend consistent with findings in related studies~\cite{Wang:2022aga}.
To evaluate the uncertainties in our results, different values of the cutoff parameter $q_{\max}$ have been considered, as reflected in the tables.

}

\R{
\begin{itemize}
\item \textbf{Decuplet baryon}s:
\end{itemize}
}
{\color{black}
Tables~\ref{NEW3} and~\ref{NEW4} present the results for the vector meson and decuplet baryon interaction in the sectors with strangeness $S=-1$ and $S=-2$, respectively. Specifically, Table~\ref{NEW3} shows the dynamically generated state from the $\bar{D}^* \Sigma^*$ interaction with isospin $I=1/2$, while Table~\ref{NEW4} corresponds to the $\bar{D}^* \Xi^*$ system with $I=0$.
}

\begin{table}[H]
\centering
\caption{The pole (in MeV) and its coupling constants $g_i$ and $g_iG_i$ (in MeV) to the channels in the $I=1/2,~S=-1$ sector. }
\label{NEW3}
\setlength{\tabcolsep}{28pt}
\begin{tabular}{ll|lc}
\hline \hline
$q_\text{max}$ &     Pole          &       ~            &  $\bar{D}^* \Sigma^*$ (Th = 3391 MeV) \\
\hline
{550 } & $\textbf{3386 }$  & $g_i$&  $2.03  $   \\
   &              & $g_iG_i$ & $-17.46   $       \\
\cline{3-4}
{600 } & $\textbf{3381  }$  & $g_i$&  $2.52  $   \\
   &              & $g_iG_i$ & $-21.76   $       \\
\cline{3-4}
{630 } & $\textbf{3377 }$  & $g_i$&  $2.81 $   \\
   &              & $g_iG_i$ & $-24.27  $ \\
\cline{3-4}
{650 } & $\textbf{3374 }$  & $g_i$&  $3.00  $   \\
   &              & $g_iG_i$ & $-25.93   $       \\
\hline
\hline
\end{tabular}
\end{table}

\begin{table}[H]
\centering
\caption{The pole (in MeV) and its coupling constants $g_i$ and $g_iG_i$ (in MeV) to the channels in the $I=0,~S=-2$ sector.}
\label{NEW4}
\setlength{\tabcolsep}{28pt}
\begin{tabular}{ll|lc}
\hline \hline
$q_\text{max}$ &     Pole          &       ~            &  $\bar{D}^* \Xi^*$ (Th = 3540.36 MeV) \\
\hline
{550 } & $\textbf{3533 }$  & $g_i$&  $2.18   $   \\
   &              & $g_iG_i$ & $-18.83    $       \\
\cline{3-4}
{600 } & $\textbf{3527 }$  & $g_i$&  $2.66   $   \\
   &              & $g_iG_i$ & $ -22.96   $       \\
\cline{3-4}
{630 } & $\textbf{3523  }$  & $g_i$&  $2.94   $   \\
   &              & $g_iG_i$ & $ -25.41   $       \\
\cline{3-4}
{650 } & $\textbf{3519}$  & $g_i$&  $ 3.13  $   \\
   &              & $g_iG_i$ & $-27.03    $       \\
\hline
\hline
\end{tabular}
\end{table}
{\color{black}
In both sectors, we observe the generation of bound states below the respective thresholds of $\bar{D}^* \Sigma^*$ and $\bar{D}^* \Xi^*$, with the binding becoming slightly stronger as the cutoff parameter $q_{\text{max}}$ increases. The strength of the coupling constants $g_i$ and the values of $g_i G_i$ indicate that these states couple strongly to the single available channel in each case. 

Compared with the octet baryon cases shown previously, the binding energies in the decuplet sectors are slightly smaller. This reflects a general trend where vector–decuplet baryon interactions still generate molecular states, though they tend to be more weakly bound than their octet counterparts.
}
\section{discussion}
\label{dis}

There is not much theoretical work done on this issue, and no experimental information so far. There are early calculations using quark models which predict, using color spin hyperfine interaction~\cite{Lipkin:1987sk} and chromomagnetic interaction~\cite{Gignoux:1987cn} that a $\bar csqqq$ pentaquark should be bound, although no numbers for the masses are provided.

In Ref.~\cite{Hofmann:2005sw} a formalism using SU(4) symmetry is used to make prediction for the interaction in  the coupled channels that we have studied. Bound states are found for the $S=-1,-2$ systems and the bindings are large, of the order of 300 MeV, although they vary depending on different scenarios studied. There is, however, an important difference with respect to our approach in the way the $G$ functions are regularized, where a $G'(\sqrt{s})$ function is taken as   $G'(\sqrt{s})=G(\sqrt{s})-G(\mu)$. This prescription gives positive values of $G'$ for $\sqrt{s}<\mu$, which can lead to very bound states with a repulsive potential (see discussions in Ref.~\cite{Wu:2010rv} to this respect).

A different approach is followed in Ref.~\cite{Yan:2023ttx} where the interaction is taken from meson exchange studied with single channels. While we discuss below that coupled channels are important in this problem, Ref~\cite{Yan:2023ttx} shows that there is indeed attraction in the $\bar D\Sigma$,  $\bar D^*\Sigma$, and $\bar D\Xi$,  $\bar D^*\Xi$ channels, as we also find here.

In Ref.~\cite{Yalikun:2021dpk} the relevance of coupled channels is shown and the  authors consider the $\bar{D}_s N$, $\bar{D} \Lambda$,  $\bar{D} \Sigma$, $\bar{D}^* \Lambda$,  $\bar{D}^* \Sigma$. The  $\bar{D}^*_s N$  channel is not considered and the $S=-2$ sector is not studied. The coupled channels provide more attraction than when considering $\bar{D} \Sigma$,  $\bar{D}^* \Sigma$  single channels. The results in~\cite{Yalikun:2021dpk} depend on two unknown parameters, $a$, $\Lambda$, where $\Lambda$ is a cutoff appearing in a form factor. Yet, even playing with uncertainties the binding energies found in~\cite{Yalikun:2021dpk} are smaller than found here. In our work we take two independent blocks for $\bar D_{s} B$ and $\bar D^*_{s} B$. They can mix through the exchange of pseudoscalar mesons, but, as shown in~\cite{Dias:2021upl}, these interactions have little influence on the binding energies, when one has contribution from vector exchange, although they provide decay widths for the $\bar D^*_{s} B$  states, which are found small in~\cite{Yalikun:2021dpk}. In~\cite{Yalikun:2021dpk} the authors also consider $\sigma$  exchange. From  the  point of view of chiral dynamics, the $\sigma$ is a resonance obtained from the $\pi\pi$ interaction~\cite{Oller:1997ti,Kaiser:1998fi} and in Ref.~\cite{Oset:2000gn} the $\sigma$ exchange in the $NN$ interaction is studied from this perspective. Yet, when this approach is used for the interaction in problems, involving $D$ mesons similar to the present one, it {renders} very  small contributions~\cite{Aceti:2014kja,Aceti:2014uea}. We see the source of our larger bindings in the form factors used in~\cite{Yalikun:2021dpk}, $(\Lambda^2-m_\text{ex}^2)/(\Lambda^2-q^2)$. While this form factor is often used, it has the effect of reducing the strength of the vector exchange, which is the main source of the interaction. The factor $\Lambda^2-m_\text{ex}^2$   in the numerator of this form factor does not appear in the   local hidden gauge approach, or the equivalent chiral Lagrangians in the SU(3) sector, which are derived from different principles (see equivalence in Ref.~\cite{Dias:2021upl}). For $\omega$ exchange the factor $(\Lambda^2-m_\text{ex}^2)/\Lambda^2$ gives a reduction of a factor 0.57 in the vector exchange. There is more to it . In table~\ref{coeff_S1} we see that there is a transition potential from the  $\bar{D} \Sigma$  to the  $\bar{D}_s N$  channel. This transition is very important, because, as shown in~\cite{ Hyodo:2013nka,Wang:2022pin,Aceti:2014ala}, if one has two coupled channels, 1, 2, one can eliminate one channel, say 2, by considering an effective potential in the remaining channel,
$$
V_\text{eff}= V_{11} + \frac{V_{12}^2G_2}{1-V_{22}G_2}\sim V_{11}+ V_{12}^2G_2.
$$
Since $V_{12}$ appears now {quadratic},
one can see that the factor $\Lambda^2-m_\text{ex}^2$ would reduce the contribution of the coupled channel by about a factor 0.33. To strengthen the importance of the coupled channels, it is worth recalling that in the study of the $P_c$ states the channel $\bar{D}^{(*)} \Sigma_c$  does not couple to $\bar{D}^{(*)} \Lambda_c$~\cite{Xiao:2013yca} but in the study of the $P_{cs}$ states, the  $\bar{D}_s^{(*)} \Lambda_c$  couples strongly to $\bar D \Xi_c$  and this is responsible for a much larger bindings in the $P_{cs}$ states than in the $P_{c}$ ones~\cite{Feijoo:2022rxf}. The former discussion can justify why we get more binding than in Ref.~\cite{Yalikun:2021dpk}. With the values of $q_\text{max}$ obtained from our study of the $P_{cs}$ and $\Omega_c$ states with the same formalism, we think that the predicted values are realistic. However,it is clear that having some of the predicted states measured in some experiment would produce a valuable information to tune the parameters of the theory and be more accurate in the other predictions made. 

\section{Conclusion}
\label{conclusion}
In this work, we studied meson-baryon interactions involving anticharm and strangeness within the framework of the extended local hidden gauge approach. Using vector meson exchange as the driving mechanism, we constructed the interaction potentials and solved the coupled-channel Bethe-Salpeter equation to explore the possible formation of molecular states in the strangeness sectors $S=-1$, $S=-2$, $S=-3$, and $S=-4$.

In the $S=-1$ sector, we considered the coupled channels $\bar{D}_s N$, $\bar{D} \Lambda$, and $\bar{D} \Sigma$. The interaction among them is attractive, and a bound state is dynamically generated with a mass around 2888 MeV (18 MeV binding with respect to the $\bar{D}_s N$ threshold). This state couples most strongly to the $\bar{D} \Sigma$ channel, but the wave function at the origin is largest for the $\bar{D}_s N$ channel, indicating its dominant molecular component.
{\color{black} We have analyzed both $I=1/2$ and $I=3/2$ configurations in this sector. The bound state appears only in the $I=1/2$ case, while the $I=3/2$ interaction is repulsive and does not generate any state, unless we use large values of $q_\text{max}$ as a consequence of the attraction induced by the coupled channels.}

In the $S=-2$ sector, the coupled channels are $\bar{D}_s \Lambda$ and $\bar{D} \Xi$. An attractive interaction leads to the formation of another bound state with mass around 3057 MeV (27 MeV binding with respect to the $\bar{D}_s \Lambda$ threshold). The coupling analysis shows that this state is dominated by the $\bar{D} \Xi$ component.
{\color{black} For this sector, we explored both $I=0$ and $I=1$ configurations. A bound state is found in the $I=0$ case, while the $I=1$ configuration does not support the formation of any bound or resonant state.}

In contrast, in the $S=-3$ sector composed of the $\bar{D}_s \Xi$ channel and the $S=-4$ sector composed of the $\bar{D}_s \Omega$ channel, the interactions are repulsive and do not support bound or resonant states. Consequently, no poles are found in these sectors.
{\color{black} In particular, we examined the $S=-3$ system with isospin $I=1/2$, and the repulsive interaction prevents the generation of any state in this configuration.}

{\color{black}We have also performed calculations for systems involving both the baryon octet and the baryon decuplet, considering different isospin configurations to explore possible state formations more comprehensively. An uncertainty analysis was carried out by varying the momentum cutoff $q_{\text{max}}$, which allows us to estimate the model dependence of the results.}

\R{
We have also extended our study to systems where the pseudoscalar anticharmed mesons ( $\bar{D}$ and $\bar{D}_s$) are replaced by their vector counterparts ($\bar{D}^*$ and $\bar{D}_s^*$). The interaction potentials remain essentially unchanged, except for the corresponding mass differences. In these vector-baryon systems, we generally observe slightly stronger binding compared to the pseudoscalar cases, resulting in molecular states that are more deeply bound across different strangeness sectors. The dominant components in these states are analogous to those found in the pseudoscalar-baryon systems, reflecting a similar underlying dynamics. This tendency for vector meson interactions to produce somewhat deeper bound states is a commonly reported feature in related theoretical approaches.
}

These findings suggest that baryon molecular states are more likely to form in systems containing one or two strange quarks, while states with three or four strange quarks do not favor such structures due to a repulsive potential. This strangeness dependence of the interaction dynamics provides useful insights into the nature of exotic baryons and helps guide future experimental searches.

The predicted bound states in the $S=-1$ and $S=-2$ sectors are promising candidates for detection at facilities such as LHCb, Belle II, or BESIII, where such states can be produced and analyzed.

In summary, our study demonstrates that the local hidden gauge formalism, combined with coupled-channel dynamics is a powerful tool for predicting novel hadronic molecules in the heavy-flavor sector. The inclusion of wave function analysis, and extension to vector meson interactions provides a comprehensive understanding of the internal structure of these states. {\color{black}The consideration of both octet and decuplet baryons, along with different isospin states and cutoff dependence, further enhances the robustness and completeness of the analysis.} Our results offer valuable guidance for both theoretical and experimental investigations of exotic states.

\section*{Appendix: Evaluation of the $VBB$ vertices with wave functions}\label{app}
 In Ref.~\cite{Debastiani:2017ewu}, it was shown that the chiral Lagrangians, or the contribution of vector exchange in the local hidden gauge approach, could be evaluated using the explicit quark wave functions of mesons and baryons.

\begin{figure}[H]
  \centering
  \includegraphics[width=4.5cm]{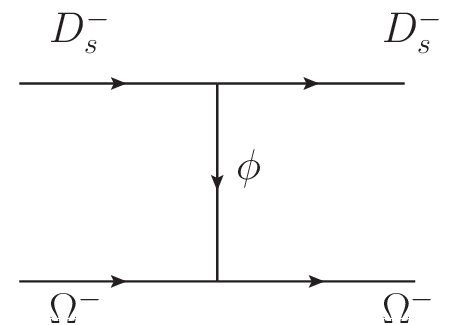}
  \caption{Diagram for $D_s^-\Omega \to D_s^-\Omega$   interaction through $\phi$ exchange.}
  \label{app_fig2}
\end{figure}

\begin{figure}[H]
  \centering
  \includegraphics[width=4.5cm]{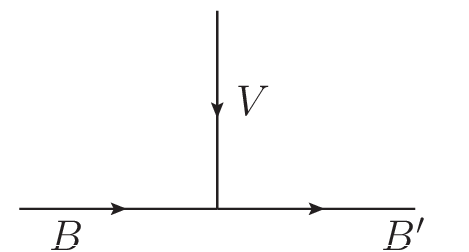}
  \caption{Diagram for the $VBB$ vertex.}
  \label{app_fig1}
\end{figure}

In the  $D_s^-\Omega \to D_s^-\Omega$ interaction, we have the contribution show in Fig.~\ref{app_fig2},
and we need to evaluate the lower vertex, corresponding to  Fig.~\ref{app_fig1}.
It is clear that one can only exchange a $\phi$ meson in Fig.~\ref{app_fig2}. The Lagrangian for $\rho^0,~\omega,~ \phi$    couplings to $B,~B$ is given by
\begin{equation}
   {{\cal L}}_{VBB} \equiv g\Bigg\langle~B'~ \Bigg| \left\{ 
         \begin{aligned}
           & \frac{1}{\sqrt{2}} (u\bar{u} - d\bar{d}),\quad \rho^0 \\
           & \frac{1}{\sqrt{2}} (u\bar{u} + d\bar{d}),\quad \omega \\
           & \qquad~~ s\bar{s}, \qquad\quad~\, \phi \\
         \end{aligned} \right\}\Bigg|~B~\Bigg\rangle~ \epsilon^0,
     \label{eq:L_VBB}
\end{equation}
where we have already taken $\gamma^\mu\equiv\gamma^0\equiv1$, with $\epsilon^\mu$ the vector polarization, and hence, there is no spin dependence, and $B,~B'$, are the baryon wave functions written in terms of quarks~\cite{Close:1979bt}. Hence, in this case we have
\begin{equation}
  {{\cal L}}_{\phi\Omega\Omega} \equiv g\langle sss~ \chi_s|~  s\bar s~|sss ~\chi_s\rangle=3g,
     \label{eq:L_phiOO}
\end{equation}
where $\chi_s$ is the spin symmetric wave function.

The upper vertex $D_sD_s\phi$   in Fig.~\ref{app_fig2}, is readily evaluated from Eq.~(\ref{lPPV}) and gives 
\begin{equation}
-it_{\phi D_s^-D_s^-} =ig(k^0 + k^{\prime 0})\epsilon^0.
\end{equation}
Thus, the $D_s^-\Omega \to D_s^-\Omega$ potential is given by
\begin{equation}
-iV =ig(k^0 + k^{\prime 0})\frac{i}{q^2-M_V^2}(-g^{00})\,i\,3\,g,
\end{equation}
from where, neglecting $q^2$ versus $M_V^2$, one obtains
\begin{equation}
V = C \frac{1}{4f_\pi^2}(k^0 + k^{\prime 0});\qquad C=3.
\end{equation}

For other cases we proceed equally, talking the wave functions $B,~B'$   of the quark models as fully symmetric flavor wave functions~\cite{Close}. The spin symmetric wave function does not play a role because the vertex is spin independent in the approximations done.

\section*{ACKNOWLEDGMENTS}
We would like to thank Prof. Maojun Yan for useful comments.
This work is partly supported by the National Natural Science
Foundation of China under Grants  No. 12405089 and No. 12247108 and
the China Postdoctoral Science Foundation under Grant
No. 2022M720360 and No. 2022M720359.  This work is also supported by the Spanish
Ministerio de Economia y Competitividad (MINECO) and European FEDER 
funds under Contracts No. FIS2017-84038-
C2-1-P B, PID2020- 112777GB-I00, and by Generalitat Valenciana under 
con- tract PROMETEO/2020/023. This project
has received funding from the European Union Horizon 2020 research and 
innovation programme under the program
H2020- INFRAIA-2018-1, grant agreement No. 824093 of the STRONG-2020 
project. This work is supported by the Spanish Ministerio de Ciencia e 
Innovacion (MICINN) under contracts PID2020-112777GB-I00, 
PID2023-147458NB- ´
C21 and CEX2023-001292-S; by Generalitat Valenciana under contracts 
PROMETEO/2020/023 and CIPROM/2023/59.

\bibliography{refs.bib} 
\newpage

\end{document}